\documentclass[twocolumn,showpacs,preprintnumbers,amsmath,amssymb,letterpaper,prd]{revtex4}

\begin{document}
\sloppy

\title{Kramers-Moyall cumulant expansion for the probability distribution of parallel transporters in quantum gauge fields}
\author{P. V. Buividovich}
\email{buividovich@tut.by}
\affiliation{Belarusian State University, 220080 Belarus, Minsk, Nezalezhnasti av. 4}
\author{V. I. Kuvshinov}%
\email{v.kuvshinov@sosny.bas-net.by}
\affiliation{JIPNR, National Academy of Science, 220109 Belarus, Minsk, Acad. Krasin str. 99}
\date{8 April 2006}

\begin{abstract}
 A general equation for the probability distribution of parallel transporters on the gauge group manifold is derived using the cumulant expansion theorem. This equation is shown to have a general form known as the Kramers-Moyall cumulant expansion in the theory of random walks, the coefficients of the expansion being directly related to nonperturbative cumulants of the shifted curvature tensor. In the limit of a gaussian-dominated QCD vacuum the obtained equation reduces to the well-known heat kernel equation on the group manifold.
\end{abstract}
\maketitle

\section{Introduction}
\label{sec:Intro}

 One of the most popular test objects used to detect quark confinement is a Wilson loop calculated in some representation of the gauge group. Measurements of the Wilson loops allow one to investigate the interaction potential between static quarks in different representations of the gauge group. For instance, Wilson area law corresponds to QCD string with constant tension \cite{Wilson:74}. Most numerical simulations indicate that Wilson loops indeed obey the Wilson area law at intermediate distances $\sim 0.2 \ldots 1$ fm \cite{Bali:00, Deldar:00}. However, at large distances QCD string may be torn apart by a quark-antiquark pair born from the vacuum, or the charge of quarks may be screened by dynamical charges, and as a result Wilson area law is violated. Another distinct feature of QCD vacuum is the Casimir scaling phenomenon, which was observed in lattice simulation \cite{Bali:00, Deldar:00}. Violation of Casimir scaling is also related to screening of static sources. Screening effects in $SU(2)$ Yang-Mills theory were detected in numerical simulations \cite{Forcrand:00, Kallio:02}. A proper theoretical description of screening in QCD vacuum is still absent, although many beautiful phenomenological models were proposed (for a review see, for example, \cite{Shevchenko:00, Simonov:96}).

 An interesting phenomenological description of screening effects, which is based on the mathematical analogy between loop dynamics and the motion of a random walker on the gauge group manifold, was proposed recently in \cite{deHaro:04, Brzoska:05, Buividovich:06:2, Arcioni:05}. The theory of random walks on Lie groups is an interesting mathematical subject \cite{Varopoulos:94, Varopoulos:96}, and one can expect that physical applications of the results obtained therein may provide important insights into the physics of confinement. For example, an important results concerning the connection between Chern-Simons theory, B-model branes on noncommutative two-sphere and random walks in the fundamental Weyl chamber of Lie algebras were obtained in \cite{deHaro:04}. In the work \cite{Brzoska:05} it was observed that Casimir scaling corresponds to a free diffusion on the group manifold, which is described by the heat kernel equation. This could be expected, since it is known from lattice gauge theory that infrared behavior of Yang-Mills theories in higher dimensions can be described by two-dimensional Yang-Mills theory (dimensional reduction scenario, \cite{Makeenko:84}), and the partition function of two-dimensional Yang-Mills theory obeys the heat kernel equation \cite{Migdal:76}. In order to describe screening effects one should somehow modify the simplest heat kernel equation. A straightforward modification is to add a drift term to the equation, which was interpreted as the Fokker-Planck equation on the group manifold in \cite{Brzoska:05, Arcioni:05}. Such a modification led to many nontrivial physical consequences, such as the emergence of confining $k$-strings \cite{Arcioni:05}. $Z_{N}$ symmetry of the group center was found to play a crucial role in transition from Casimir scaling to screening regime \cite{Brzoska:05, Arcioni:05}, which has a nice interpretation in terms of central vortices. A non-trivial but solvable modification of QCD$_{2}$ matrix model which describes screening effects was discovered in \cite{Arcioni:05}. In our work \cite{Buividovich:06:2} we have pointed out another possibility to generalize the heat kernel equation, namely to include the terms with higher-order derivatives with respect to group coordinates. An advantage of such a generalization is that there is no need to introduce some external force in the Fokker-Planck equation "by hands", as it is done in the works \cite{Brzoska:05, Arcioni:05}. Thus geometric properties of the group manifold (homogeneity and absolute parallelism) are preserved and the symmetry of the equation under group transformations is explicit.

 The aim of this paper is to provide a theoretical ground for the phenomenological analysis of \cite{Brzoska:05, Buividovich:06:2, Arcioni:05} and to derive the most general differential equation for the probability distribution of parallel transporters on the group manifold from the first principles. The classical loop equation \cite{PolyakovGaugeStrings} is first rewritten as a flow equation on the group manifold. Then we apply the cumulant expansion theorem to this flow equation and find that the probability distribution obeys the equation which can be obtained from the heat kernel equation by adding higher-order differential operators. This equation generalizes the Fokker-Planck equation used in \cite{Brzoska:05, Arcioni:05} and is known as the Kramers-Moyall cumulant expansion in the theory of random walks \cite{RiskenFokkerPlanck}. The coefficients in the obtained equation can be expressed in terms of nonperturbative cumulants of the shifted curvature tensor, which are the basic objects in the method of field correlators \cite{Dosch:02, Simonov:96, Steffen:03}. In the limit of a gaussian-dominated QCD vacuum, where only the second-order cumulant survives, this equation reduces to the heat kernel equation, in accordance with the results of \cite{Brzoska:05, Buividovich:06:2, Arcioni:05}. We suppose that the proposed generalization of the heat kernel equation may be helpful in finding effective action for QCD string which takes into account screening effects. Kramers-Moyall cumulant expansion also emerges naturally as a continuum approximation of discrete random walks, which can be interpreted as a continuum limit of a lattice gauge theory.

 The paper is organized as follows: in this section we introduce the loop variables and the classical loop equations \cite{PolyakovGaugeStrings}. Since we extensively use geometric constructions on the group manifold throughout the paper, in the next section we make a brief review of geometric properties of compact simple groups and introduce the group laplacian. In the section \ref{sec:ClassicalLoop} the probability distribution of parallel transporters on the group manifold is introduced and its classical limit is considered. Finally, in the section \ref{sec:CumulantExpansion} we apply the cumulant expansion theorem to obtain the main result of this paper, namely the differential equation for the probability distribution of parallel transporters on the group manifold. 

 We consider gauge fields $\hat{A}_{\mu} (x)$ in some representation of the Lie algebra of the gauge group $G$, which is supposed to be compact and simple. To any smooth closed path $\gamma$ one can attribute parallel transport operator which can be written as a path-ordered exponent:
\begin{eqnarray}
\label{ParallelTransportOperatorDefinition}
    \hat{\Psi} \left( \gamma \right) = \mathcal{P} \exp \left( i \int \limits_{\gamma} dx^{\mu} \hat{A}_{\mu} \right)
\end{eqnarray}   
Parallel transport operator is the element of some representation of the group $G$, therefore in classical theory to each loop $\gamma$ corresponds some element $g_{0}(\gamma)$ of the gauge group $G$. In order to calculate the path-ordered exponent in (\ref{ParallelTransportOperatorDefinition}), one should choose some initial point $x_{0}$ on the loop $\gamma$. Parallel transport operator transforms under gauge transformations in the following way:
\begin{eqnarray}
\label{ParallelTransporterGaugeTransforms}
\hat{\Psi} \left( \gamma \right) \rightarrow \hat{T} \left( f(x_{0}) \right) \hat{\Psi} \left( \gamma \right) \hat{T} \left( f^{-1}(x_{0}) \right) \nonumber \\
g_{0}(\gamma) \rightarrow f(x_{0}) g_{0}(\gamma) f^{-1}(x_{0})
\end{eqnarray}
Thus gauge orbits for loop variables are group classes, i.e. the points which belong to the same group class should be identified, and points in the physical phase space are actually the group classes.

We consider a one-parameter family of loops $\gamma(\tau)$, assuming that at fixed $\tau$ the loop $\gamma(\tau)$ is parametrized with some variable $\sigma$, i.e. $\gamma(\tau): \sigma \rightarrow x^{\mu}(\sigma, \tau), \quad x^{\mu}(0) = x^{\mu}(2 \pi)$. Group element which corresponds to the loop $\gamma(\tau)$ in classical theory will be denoted as $g_{0}(\tau) \equiv g_{0}\left(\gamma(\tau) \right)$. In order to impose boundary conditions in subsequent equations we suppose that $\tau = 0$ corresponds to a loop of zero area, for which $g_{0}\left(\gamma(0)\right) = 1$. We will also assume that the worldsheet swept by the loops $\gamma(\tau)$ is simply connected. This assumption allows us to introduce the shifted curvature tensor $\tilde{F}_{\mu \nu}$ on this worldsheet, as is done in the proof of the non-Abelian Stokes theorem \cite{Diosi:83, Fishbane:81}:
\begin{eqnarray}
\label{ShiftedCurvatureTensor}
 \tilde{F}_{\mu \nu} (x) = \hat{U}(x_{0}, x, \chi) \hat{F}_{\mu \nu} (x) \hat{U}(x, x_{0}, \chi)
\end{eqnarray}
where $\hat{F}_{\mu \nu} = \partial_{[\mu} \hat{A}_{\nu]} - i [\hat{A}_{\mu},\hat{A}_{\nu}]$ is the curvature tensor of Yang-Mills field, $\hat{U}(x, y, \chi) = \mathcal{P} \exp \left( i \int \limits_{y; \: \chi}^{x} dx^{\mu} \hat{A}_{\mu} \right)$ is the non-Abelian phase factor along the path $\chi$, $x_{0}$ is the reference point on the surface and $\chi$ is some path which connects the reference point $x_{0}$ and the point $x$. Shifted curvature tensor is the curvature tensor parallel transported from the point $x$ to the reference point $x_{0}$ along the path $\chi$.
 
 We start with the Polyakov-Migdal variational equation for loop variables in classical Yang-Mills field (\cite{PolyakovGaugeStrings}, chapter 7):
\begin{eqnarray}
\label{PolyakovEquation}
 \frac{\delta \hat{\Psi} (\gamma) }{\delta x^{\mu} (\sigma)} \: \hat{\Psi}^{\dag}(\gamma) = i \tilde{F}_{\mu \nu} \left(x (\sigma) \right) \frac{d x^{\nu}}{d \sigma}
\end{eqnarray}
For our continuous set of loops the equation (\ref{PolyakovEquation}) reduces to a linear differential equation:
\begin{eqnarray}
\label{PolyakovEquationOneParametric}
\begin{array}{l}
 \frac{d \hat{\Psi} (\tau)}{d \tau} \: \hat{\Psi}^{\dag}(\tau) = \int \limits_{\gamma(\tau)} d \sigma \: \partial_{\tau} x^{\mu}(\sigma, \tau) \frac{\delta \hat{\Psi} \left(\gamma(\tau) \right) }{\delta x^{\mu} (\sigma, \tau)} \hat{\Psi}^{\dag}\left(\gamma(\tau) \right) = \\ = i \int \limits_{\gamma(\tau)} d \sigma \partial_{\sigma} x^{\nu} (\sigma, \tau) \partial_{\tau} x^{\mu}(\sigma, \tau) \tilde{F}_{\mu \nu} \left(x (\sigma, \tau) \right)
\end{array}
\end{eqnarray}
where $\hat{\Psi} (\tau) \equiv \hat{\Psi} \left(\gamma (\tau) \right)$ and $\partial_{\sigma} = \frac{\partial}{\partial \sigma}$, $\partial_{\tau} = \frac{\partial}{\partial \tau}$. Boundary condition for this equation is $\hat{\Psi} (0) = \hat{I}$, because we have assumed that $g_{0}(0) = 1$. 

\section{Geometry of the group manifold}
\label{sec:GroupGeometry}

 In order to proceed with the analysis of the equation (\ref{PolyakovEquationOneParametric}) it is necessary to introduce some geometric constructions on the group manifold. In this section we make a brief review of geometric properties of simple compact groups. In order to relate our notations to those of another authors \cite{Brzoska:05, Arcioni:05} we also introduce here the group laplacian. 

 Group generators in the fundamental representation $\hat{T}_{a}$, $\hat{T}^{\dag}_{a} = \hat{T}_{a}$ are chosen to satisfy the normalization ${\rm Tr} \: \hat{T}_{a} \hat{T}_{b} = \delta_{ab}$ and obey the commutation relations $[\hat{T}_{a}, \hat{T}_{b}] = i C^{c}_{ab} \hat{T}_{c}$. Suppose that group elements are parametrized by some coordinates $g^{\alpha}$ (for the sake of brevity we will use the same letters to denote the elements of the group and their coordinates). In this paper we will work in several different spaces: the space-time where our Yang-Mills fields live, the group manifold and the Lie algebra (tangent space to the group manifold at the group identity). From now on we will use latin indices $a, b, c, \ldots$ to denote the elements of the Lie algebra, the first greek indices $\alpha, \beta, \ldots$ for tensors on the group manifold, and the indices from the middle of the greek alphabet $\mu, \nu, \ldots$ for tensor fields in space-time. We define the vector fields $L^{\alpha}_{a}(g)$ and $R^{\alpha}_{a}$ which are induced by the left and right action of the group generators respectively \cite{DeWittGroupsAndFields, BarutRonczkaGroupRepresentation}:
\begin{eqnarray}
\label{LeftRightVectorFields}
(1 + i \epsilon^{a} \hat{T}_{a}) \hat{U}(g^{\alpha}) = \hat{U}(g^{\alpha} + L^{\alpha}_{a}(g) \epsilon^{a}) \nonumber \\
\hat{U}(g^{\alpha})(1 + i \epsilon^{a} \hat{T}_{a}) = \hat{U}(g^{\alpha} + R^{\alpha}_{a}(g) \epsilon^{a})
\end{eqnarray}
where $\hat{U}(g^{\alpha})$ are the matrices of the fundamental representation and $\epsilon^{a}$ are arbitrary infinitely small parameters. Let us also introduce the derivatives $\nabla_{a}$ and $\tilde{\nabla}_{a}$ (not to be confused with Yang-Mills covariant derivatives in space-time) along these vector fields:
\begin{eqnarray}
\label{LeftRightDerivatives}
\nabla_{a} = L^{\alpha}_{a}(g) \frac{\partial}{\partial g^{\alpha}}, \quad
\tilde{\nabla}_{a} = R^{\alpha}_{a}(g) \frac{\partial}{\partial g^{\alpha}}
\end{eqnarray}
These derivatives obey the following commutation relations:
\begin{eqnarray}
\label{LeftRightCommutationRelations}
\left[ \nabla_{a}, \nabla_{b} \right] = C^{c}_{ab} \nabla_{c}, \quad \left[ \tilde{\nabla}_{a}, \tilde{\nabla}_{b} \right] = - C^{c}_{ab} \tilde{\nabla}_{c} \nonumber \\
\left[ \tilde{\nabla}_{a}, \nabla_{b} \right] = 0
\end{eqnarray}
Thus the derivatives $\nabla_{a}$ or $\tilde{\nabla}_{a}$ acting on the functions on the group manifold build an infinite-dimensional reducible representation of the Lie algebra of the group $G$. Corresponding representation of the group $G$ is the regular representation, which contains all finite-dimensional irreducible unitary representations. Quadratic Casimir operator of this representation is \cite{Berezin:62}:
\begin{eqnarray}
\label{LaplacianVectorFields}
   \Delta = \nabla_{a} \nabla_{a} = \tilde{\nabla}_{a} \tilde{\nabla}_{a}
\end{eqnarray}
Eigenvalues of $ - \Delta$ are the eigenvalues of the quadratic Casimir operators in irreducible unitary representations. The minus sign arises here because the derivatives $\nabla_{a}$ are anti-hermitean, unlike the hermitean generators $\hat{T}_{a}$.

While the derivatives $\nabla_{a}$ and $\tilde{\nabla}_{a}$ generate left and right group multiplications respectively, their difference generates shifts within group classes:
\begin{eqnarray}
\label{ShiftsWithinGroupClasses}
\nabla_{a} \hat{U}(g) = i \hat{T}_{a} \hat{U}(g), \quad \tilde{\nabla}_{a} \hat{U}(g) = i \hat{U}(g) \hat{T}_{a} \nonumber \\
\left(\nabla_{a} - \tilde{\nabla}_{a} \right) \hat{U}(g) = i \left[ \hat{T}_{a}, \hat{U}(g) \right]
\end{eqnarray}
In this paper we will work in a Hilbert space of functions on group classes. Such functions obey the constraint $\phi(g) = \phi(f g f^{-1})$ for any group element $f$. An infinitesimal version of this constraint is:
\begin{eqnarray}
\label{FunctionOnClasses}
\left(\nabla_{a} - \tilde{\nabla}_{a} \right) \phi(g) = 0
\end{eqnarray}
Thus the Hilbert space of functions on group classes is the subspace of the full Hilbert space of functions on the group manifold, spanned on the solutions of the linear equation (\ref{FunctionOnClasses}).

 An invariant metric on the group manifold is fixed up to a constant by requiring infinitesimal distances to be invariant under group multiplication \cite{DeWittGroupsAndFields}. This means that the vectors $L^{a}_{\alpha}$ are Killing vectors, which immediately gives for the group metric (Killing form) $h_{\alpha \beta}$:
\begin{eqnarray}
\label{Metrics}
 h^{\alpha \beta}(g) = L^{\alpha}_{a}(g) L^{\beta}_{a}(g), \quad h^{\alpha \beta} h_{\beta \gamma} = \delta^{\alpha}_{\gamma}
\end{eqnarray}
An invariant measure on the group manifold (the Haar measure) is $d \mu(g) = \sqrt{\det{h_{\alpha \beta}(g)}} \prod \limits_{\gamma} dg^{\gamma}$.

 Group structure of the manifold allows also to define the parallel transport and the  connection on it. Parallel transport is defined as follows: consider two infinitely close points with coordinates $g^{\alpha}$ and $g^{\alpha} + d g^{\alpha}$. Suppose that under left group multiplication these points transform into $\tilde{g}^{\alpha}$ and $\tilde{g}^{\alpha} + d \tilde{g}^{\alpha}$. Then $d \tilde{g}^{\alpha}$ is said to be the vector $d g^{\alpha}$ parallel transported from the point $g$ to the point $\tilde{g}$ \cite{DeWittGroupsAndFields}. The corresponding connection is $\Gamma^{\gamma}_{\alpha \beta} = L^{\gamma}_{a} \frac{\partial}{\partial g^{\alpha}} L_{\beta \: a}$, where indices are lowered and raised with the group metrics $h_{\alpha \beta}$. One uses this connection to define a covariant derivative $\nabla_{\alpha}$ on the group manifold. Group manifold is a space of absolute parallelism, i.e. the space with zero curvature but nonzero torsion: $[\nabla_{\alpha}, \nabla_{\beta}] = S^{\gamma}_{\alpha \beta} \nabla_{\gamma}$. We use the covariant derivatives $\nabla_{\alpha}$ to redefine $\Delta$ and to extend its action on arbitrary tensor fields on the group manifold:
\begin{eqnarray}
\label{Laplacian}
    \Delta = h^{\alpha \beta} \nabla_{\alpha} \nabla_{\beta}
\end{eqnarray}
For scalar functions the definitions (\ref{LaplacianVectorFields}) and (\ref{Laplacian}) are equivalent, as in this case $\Delta$ is the Beltrami-Laplace operator for the group metrics $h_{\alpha \beta}$:
\begin{eqnarray}
\label{BeltramiLaplace}
 \Delta = (\det{h_{\alpha \beta}})^{-1/2} \frac{\partial}{\partial g^{\alpha}} \left((\det{h_{\alpha \beta}})^{1/2} h^{\alpha \beta} \frac{\partial}{\partial g^{\beta}} \right)
\end{eqnarray}
Definitions (\ref{Laplacian}) and (\ref{BeltramiLaplace}) are different in the case of arbitrary geometry, but in the particular case of manifolds with group structure they coincide.

\section{Probability distribution of parallel transporters on the group manifold and its classical limit}
\label{sec:ClassicalLoop}

 In quantum theory one can not establish any deterministic correspondence between loops in space-time and points on the group manifold. Instead one can consider the probability distribution $p(g, \tau)$ which will determine the probability $dP(g, \tau) = p(g, \tau) d \mu(g)$ for the group element $g(\tau)$ associated with the loop $\gamma(\tau)$ to be within an infinitesimal volume $d \mu(g)$ on the group manifold. As the physical phase space contains only group classes, the points which belong to the same group class should be identified, which means that the physical probability distribution is actually defined on group classes. Thus we require the function $p(g, \tau)$ to be the function on group classes, i.e. to obey the constraint (\ref{FunctionOnClasses}). As will be shown further, the constraint (\ref{FunctionOnClasses}) is consistent with the final equation for $p(g, \tau)$. A full orthonormal basis in the Hilbert space of functions on group classes is built by the characters of irreducible unitary representations $\chi_{k}(g)$, where the index $k$ labels irreducible unitary representations of the group $G$, including the trivial one. An identity operator on this space is the delta-function on group classes:
\begin{eqnarray}
\label{DeltaFunctionGroupClasses}
\delta(f,g) = \sum \limits_{k} \chi_{k} (f) \bar{\chi}_{k}(g)
\end{eqnarray}
By definition, for any function on group classes $\phi(g)$ one has:
\begin{eqnarray}
\label{GroupDeltaFunctionDefinition}
 \int \limits_{G} d \mu (g) \phi(g) \delta \left(f, g \right) = \phi(f)
\end{eqnarray}

 As usual in quantum theory, we introduce the probability distribution $p(g, \tau)$ as the expectation value of the delta-function on the physical phase space \cite{Buividovich:06:2, Brzoska:05, Arcioni:05}:
\begin{eqnarray}
\label{AmplitudeDefinition}
    p(g, \tau) = \langle \: \delta \left(g, g_{0}(\tau) \right) \: \rangle
\end{eqnarray} 
where by $\langle \: \ldots \: \rangle$ we denote the vacuum expectation value. As $\langle 1 \rangle = 1$, the probability distribution (\ref{AmplitudeDefinition}) is automatically normalized to unity: $\int \limits_{G} d \mu(g) p(g, \tau) = \langle \:  \int \limits_{G} d \mu(g) \delta \left(g, g_{0}(\tau) \right) \: \rangle = 1$.

As $\delta \left(g, g_{0}(\tau) \right)$ in (\ref{AmplitudeDefinition}) is the delta-function on group classes (\ref{DeltaFunctionGroupClasses}), the function $p(g, \tau)$ is a gauge-invariant quantity. A simple calculation shows that indeed the probability distribution $p(g, \tau)$ can be expressed in terms of Wilson loops \cite{Buividovich:06:2}:
\begin{eqnarray}
\label{ProbabilityDistributionInTermsOfWilsonLoops}
 p(g, \tau) = \langle \: \delta \left(g, g_{0}(\tau) \right) \: \rangle =
\langle \: \sum \limits_{k} \chi_{k}(g)  \bar{\chi}_{k} \left( g_{0}(\tau) \right) \: \rangle = \nonumber \\ =
 \sum \limits_{k} \chi_{k}(g)  \langle \: \bar{\chi}_{k} \left( g_{0}(\tau) \right) \: \rangle = \sum \limits_{k} \chi_{k}(g) d_{k} \bar{W}_{k}\left( \gamma(\tau) \right)
\end{eqnarray}
where $W_{k}(\gamma)$ is the Wilson loop in the $k$-th representation of the gauge group and $d_{k}$ is the dimensionality of the $k$-th representation. 

 We would like to find a differential equation satisfied by $p(g, \tau)$. To this end we consider the classical limit of $p(g, \tau)$, which is given by the delta-function at the classical solution: $p_{0}(g, \tau) = \delta \left(g, g_{0}(\tau)\right)$. Differentiating $p_{0}(g, \tau)$ over $\tau$ gives:
\begin{eqnarray}
\label{ClassicalEq}
 \partial_{\tau} \: \delta \left(g, g_{0}(\tau) \right) = \frac{\partial}{\partial g_{0}^{\alpha}} \: \delta \left(g, g_{0}(\tau) \right) \: \frac{d g_{0}^{\alpha}(\tau)}{d \tau} = \nonumber \\ = - \frac{\partial}{\partial g^{\alpha}} \: \delta \left(g, g_{0}(\tau) \right) \: \frac{d g_{0}^{\alpha}(\tau)}{d \tau}
\end{eqnarray}
where we have taken into account that $\delta \left(g,f \right) = \delta \left(f,g \right)$ and hence $\frac{\partial}{\partial g^{\alpha}} \: \delta \left(g,f \right) = - \frac{\partial}{\partial f^{\alpha}} \: \delta \left(g,f \right)$. The vector field $\frac{d g_{0}^{\alpha}(\tau)}{d \tau}$ can be found from (\ref{PolyakovEquationOneParametric}):
\begin{eqnarray}
\label{VelocityOnGroup}
\frac{d g_{0}^{\alpha}(\tau)}{d \tau} = L^{\alpha}_{a}\left(g_{0}(\tau) \right) \int \limits_{\gamma(\tau)} d \sigma \partial_{\sigma} x^{\nu} \partial_{\tau} x^{\mu} \tilde{F}^{a}_{\mu \nu} \left(x (\sigma, \tau) \right)
\end{eqnarray}
It is convenient to introduce the linear differential operator $\mathcal{L}(\tau)$ which acts on functions on the group manifold as the derivative along the vector field $ - \frac{d g_{0}^{\alpha}(\tau)}{d \tau}$:
\begin{eqnarray}
\label{LinearOperatorL}
\mathcal{L}(\tau) = - \int \limits_{\gamma(\tau)} d \sigma \partial_{\sigma} x^{\nu}  \partial_{\tau} x^{\mu} \tilde{F}^{a}_{\mu \nu} \left(x (\sigma, \tau) \right) \nabla_{a}
\end{eqnarray}
Now the equation (\ref{ClassicalEq}) can be rewritten as:
\begin{eqnarray}
\label{ClassicalEqFinal}
 \partial_{\tau} p_{0}(g, \tau) = - \frac{d g_{0}^{\alpha}(g)}{d \tau} \frac{\partial}{\partial g^{\alpha}} p_{0}(g, \tau) = \mathcal{L}(\tau) p_{0}(g, \tau)
\end{eqnarray}
which is nothing but the flow equation along the vector field $\frac{d g_{0}^{\alpha}(\tau)}{d \tau}$. Initial condition for this equation is $p_{0}(g, 0) = \delta(g,1)$, as $g_{0}(0) = 1$. 

\section{Cumulant expansion}
\label{sec:CumulantExpansion}

 The probability distribution $p(g, \tau)$ is the vacuum expectation value of the function $p_{0}(g, \tau) = \delta \left(g, g_{0}(\tau)\right)$ as the functional of the gauge field $\hat{A}_{\mu}(x)$, which enters the equation for $p_{0}(g, \tau)$ through the differential operator $\mathcal{L}(\tau)$. Our aim is to obtain an equation for $p(g, \tau)$, which can be done without direct calculation of $p(g, \tau)$ by using the cumulant expansion theorem (also known as the Van-Kampen expansion) \cite{VanKampen:74, VanKampenStochasticProcesses}. Cumulant expansion was originally devised to solve linear differential equations with random coefficients. Together with the non-Abelian Stokes theorem it was successfully applied in nonperturbative QCD to find vacuum expectation values of Wilson loops \cite{Dosch:02, Simonov:96, Steffen:03}. An approach based on the cumulant expansion was further developed into the method of field correlators \cite{Dosch:02, Simonov:96, Steffen:03}. A particular case of the cumulant expansion is known in the theory of random walks as the Kramers-Moyall expansion \cite{RiskenFokkerPlanck}. 

 Applying the cumulant expansion theorem to the equation (\ref{ClassicalEqFinal}) yields the following equation for $p(g, \tau) = \langle \: p_{0}(g, \tau) \: \rangle$:
\begin{eqnarray}
\label{VanKampenExpansion}
 \partial_{\tau} p(g, \tau) = \sum \limits_{k=0}^{\infty} 
 \begin{array}{c}
 \int \ldots \int d \tau_{1} \ldots d \tau_{k} \\{\scriptstyle \tau > \tau_{1} > \tau_{2} > \ldots > \tau_{k}}
 \end{array} 
\nonumber \\
\langle \langle  \mathcal{L}(\tau) \mathcal{L}(\tau_{1}) \ldots \mathcal{L}(\tau_{k})  \rangle \rangle  p(g, \tau)
\end{eqnarray}
where double parentheses $\langle \langle \: \ldots \: \rangle \rangle$ denote the ordered cumulants \cite{VanKampen:74, VanKampenStochasticProcesses}, which extend the concept of cumulants of scalar functions onto non-commuting operators (note that the operators $\mathcal{L}(\tau)$ for different values of $\tau$ do not commute). Ordered cumulants $\langle \langle  \: \ldots \:  \rangle \rangle$ are defined implicitly by the following recurrence relation \cite{VanKampen:74, VanKampenStochasticProcesses}:
\begin{eqnarray}
\label{OrderedCumulantsRecursion}
\langle  \mathcal{L} (\tau) \mathcal{L} (\tau_{1}) \ldots \mathcal{L} (\tau_{n})  \rangle = \sum \limits_{r = 0}^{n}  \sum \limits_{
\{ i_{1} \ldots i_{r} \} \bigcup \{ j_{1} \ldots j_{n - r} \} =
\{ 1 \ldots n \} } \\
\langle \langle \mathcal{L} (\tau) \mathcal{L} (\tau_{i_{1}}) \ldots \mathcal{L} (\tau_{i_{r}})  \rangle \rangle \nonumber 
\langle  \mathcal{L} (\tau_{j_{1}}) \ldots \mathcal{L} (\tau_{j_{n - r}})  \rangle
\end{eqnarray}
where summation is performed over all decompositions of the set $\{ 1 \ldots n \}$ into two ordered subsets $\{ i_{1} \ldots i_{r} \}$ and $\{ j_{1} \ldots j_{n - r} \}$, each of which may be empty. To start the recursion we define $\langle \langle \: \mathcal{L} (\tau) \: \rangle \rangle = \langle \: \mathcal{L} (\tau) \: \rangle$. Ordered cumulants of the lowest orders are \cite{VanKampen:74}:
\begin{eqnarray}
\label{OrederedCumulantsLowestOrder}
    \langle \langle \mathcal{L} (\tau_{1}) \mathcal{L} (\tau_{2})\rangle \rangle = \langle \mathcal{L} (\tau_{1}) \mathcal{L} (\tau_{2}) \rangle - \langle  \mathcal{L} (\tau_{1})  \rangle \langle  \mathcal{L} (\tau_{2})  \rangle \nonumber \\
\langle \langle \mathcal{L} (\tau_{1}) \mathcal{L} (\tau_{2}) \mathcal{L} (\tau_{3}) \rangle \rangle = \langle  \mathcal{L} (\tau_{1}) \mathcal{L} (\tau_{2}) \mathcal{L} (\tau_{3})  \rangle  - \nonumber \\ - 
\langle  \mathcal{L} (\tau_{1}) \mathcal{L} (\tau_{2})  \rangle \langle  \mathcal{L} (\tau_{3})  \rangle - \langle  \mathcal{L} (\tau_{1})  \rangle \langle  \mathcal{L} (\tau_{2}) \mathcal{L} (\tau_{3})  \rangle - \nonumber \\ - \langle  \mathcal{L} (\tau_{1}) \mathcal{L} (\tau_{3})  \rangle \langle  \mathcal{L} (\tau_{2})  \rangle + \langle  \mathcal{L} (\tau_{1})  \rangle \langle  \mathcal{L} (\tau_{2})  \rangle \langle  \mathcal{L} (\tau_{3})  \rangle  + \nonumber \\ + \langle  \mathcal{L} (\tau_{1})  \rangle \langle  \mathcal{L} (\tau_{3})  \rangle \langle  \mathcal{L} (\tau_{2})  \rangle 
\end{eqnarray}

 Explicit substitution of $\mathcal{L}(\tau)$ in (\ref{VanKampenExpansion}) gives the desired equation for $p(g, \tau)$:
\begin{widetext}
\begin{eqnarray}
\label{GeneralFinalEquation}
\partial_{\tau} \: p(g, \tau) = \sum \limits_{k=0}^{\infty} (-1)^{k + 1} \int \limits_{\gamma(\tau)} d \sigma \partial_{\sigma} x^{\nu} \partial_{\tau} x^{\mu}
\int  \ldots \int dS^{\mu_{1} \nu_{1}} \ldots dS^{\mu_{k} \nu_{k}} \nonumber \\
\langle \langle \: \tilde{F}^{a}_{\mu \nu} \left(x (\sigma, \tau) \right) \tilde{F}^{a_{1}}_{\mu_{1} \nu_{1}} \left(x (\sigma_{1}, \tau_{1}) \right) \ldots  \tilde{F}^{a_{k}}_{\mu_{k} \nu_{k}} \left(x (\sigma_{k}, \tau_{k}) \right) \: \rangle \rangle \nabla_{a} \nabla_{a_{1}} \ldots \nabla_{a_{k}} p(g,\tau)
\end{eqnarray}
\end{widetext}
where $dS^{\mu \nu} = 1/2 \: d \sigma d \tau \: \partial_{\sigma} x^{\left[ \mu \right.} \partial_{\tau} x^{\left. \nu \right]}$ is the volume form on the worldsheet swept by the loops $\gamma(\tau)$ and integration is performed over the whole worldsheet with the restriction $\tau > \tau_{1} > \tau_{2} > \ldots > \tau_{k}$. Note that in the above equation we have used the cumulants of the shifted curvature tensor in the basis of the generators $\hat{T}_{a}$. These cumulants are defined by the following relation:
\begin{eqnarray}
\label{LieAlgebraCumulants}
 \langle \langle  \tilde{F}^{a_{1}}_{\mu_{1} \nu_{1}} \left(x (\sigma_{1}, \tau_{1}) \right) \hat{T}_{a_{1}} \ldots  \tilde{F}^{a_{k}}_{\mu_{k} \nu_{k}} \left(x (\sigma_{k}, \tau_{k}) \right)  \hat{T}_{a_{k}}  \rangle \rangle = \nonumber \\ =
\langle \langle  \tilde{F}^{a_{1}}_{\mu_{1} \nu_{1}} \left(x (\sigma_{1}, \tau_{1}) \right) \ldots  \tilde{F}^{a_{k}}_{\mu_{k} \nu_{k}} \left(x (\sigma_{k}, \tau_{k}) \right)  \rangle \rangle  \hat{T}_{a_{1}} \ldots  \hat{T}_{a_{k}}
\end{eqnarray}
where $\langle \langle \: \tilde{F}_{\mu_{1} \nu_{1}} \ldots  \tilde{F}_{\mu_{k} \nu_{k}} \: \rangle \rangle$ are the nonperturbative cumulants of the shifted curvature tensor \cite{Dosch:02, Simonov:96, Steffen:03}. Such cumulants are defined by the relation similar to (\ref{OrderedCumulantsRecursion}), where $\tau$-ordering is replaced by some surface ordering \cite{Diosi:83, Fishbane:81}, which we will assume to be the ordering with respect to $\tau$ and $\sigma$ variables. It should be noted that the shifted curvature tensor always transforms as a curvature tensor in the reference point $x_{0}$, thus the cumulants (\ref{LieAlgebraCumulants}) transform covariantly under gauge transformations. A detailed review of the properties of such cumulants can be found in \cite{Dosch:02, Simonov:96}. For instance, the lowest-order cumulants are:
\begin{eqnarray}
\label{LowestOrderCumulantsOfCurvature}
\langle \langle  \tilde{F}^{a}_{\mu_{1} \nu_{1}} \tilde{F}^{b}_{\mu_{2} \nu_{2}}  \rangle \rangle = \langle  \tilde{F}^{a}_{\mu_{1} \nu_{1}} \tilde{F}^{b}_{\mu_{2} \nu_{2}}  \rangle
 - \langle  \tilde{F}^{a}_{\mu_{1} \nu_{1}}  \rangle \langle  \tilde{F}^{b}_{\mu_{2} \nu_{2}}  \rangle \nonumber \\
\langle \langle  \tilde{F}^{a}_{\mu_{1} \nu_{1}} \tilde{F}^{b}_{\mu_{2} \nu_{2}} \tilde{F}^{c}_{\mu_{3} \nu_{3}}  \rangle \rangle = 
\langle  \tilde{F}^{a}_{\mu_{1} \nu_{1}} \tilde{F}^{b}_{\mu_{2} \nu_{2}} \tilde{F}^{c}_{\mu_{3} \nu_{3}}  \rangle - \nonumber \\ -
\langle  \tilde{F}^{a}_{\mu_{1} \nu_{1}} \tilde{F}^{b}_{\mu_{2} \nu_{2}}  \rangle  \langle  \tilde{F}^{c}_{\mu_{3} \nu_{3}}  \rangle -  
\langle  \tilde{F}^{a}_{\mu_{1} \nu_{1}}  \rangle \langle  \tilde{F}^{b}_{\mu_{2} \nu_{2}} \tilde{F}^{c}_{\mu_{3} \nu_{3}}  \rangle - \nonumber \\ -
\langle  \tilde{F}^{a}_{\mu_{1} \nu_{1}}  \rangle \langle  \tilde{F}^{b}_{\mu_{3} \nu_{3}} \tilde{F}^{c}_{\mu_{2} \nu_{2}}  \rangle + 
\langle  \tilde{F}^{a}_{\mu_{1} \nu_{1}}  \rangle \langle  \tilde{F}^{b}_{\mu_{2} \nu_{2}}  \rangle \langle  \tilde{F}^{c}_{\mu_{3} \nu_{3}}  \rangle  + \nonumber \\ + 
\langle  \tilde{F}^{a}_{\mu_{1} \nu_{1}}  \rangle \langle  \tilde{F}^{b}_{\mu_{3} \nu_{3}}  \rangle \langle  \tilde{F}^{c}_{\mu_{2} \nu_{2}}  \rangle
\end{eqnarray}
The equation (\ref{GeneralFinalEquation}) can be rewritten in a compact form as:
\begin{eqnarray}
\label{KramersMoyall}
 \partial_{\tau} p(g, \tau) = \sum \limits_{k=0}^{\infty} \eta^{a \: a_{1} \ldots  a_{k}}(\tau)  \nabla_{a} \nabla_{a_{1}} \ldots \nabla_{a_{k}} \: p(g, \tau)
\end{eqnarray}
where the coefficients $\eta^{a \: a_{1} \ldots  a_{k}}(\tau)$ are some functions of $\tau$ which can be expressed in terms of the cumulants of the shifted curvature tensor:
\begin{widetext}
\begin{eqnarray}
\label{DiffusionCoefficients}
\eta^{a \: a_{1} \ldots  a_{k}}(\tau) = (-1)^{k + 1} \int \limits_{\gamma(\tau)} d \sigma \partial_{\sigma} x^{\nu} \partial_{\tau} x^{\mu}
\begin{array}{c}
 \int \ldots \int \\{\scriptstyle \tau > \tau_{1} > \tau_{2} > \ldots > \tau_{k}}
 \end{array} dS^{\mu_{1} \nu_{1}} \ldots dS^{\mu_{k} \nu_{k}} \nonumber \\
\langle \langle \: \tilde{F}^{a}_{\mu \nu} \left(x (\sigma, \tau) \right) \tilde{F}^{a_{1}}_{\mu_{1} \nu_{1}} \left(x (\sigma_{1}, \tau_{1}) \right) \ldots  \tilde{F}^{a_{k}}_{\mu_{k} \nu_{k}} \left(x (\sigma_{k}, \tau_{k}) \right) \: \rangle \rangle
\end{eqnarray}
\end{widetext}
The equation (\ref{GeneralFinalEquation}) in the form (\ref{KramersMoyall}) has a general form of the Kramers-Moyall cumulant expansion known in the theory of random walks \cite{RiskenFokkerPlanck}.

 Gauge invariance imposes rather strict constraints on the tensor structure of the coefficients $\eta^{a \: a_{1} \ldots  a_{k}}(\tau)$. Only colour singlets can have non-zero vacuum expectation value, which means that each monomial of the form $\eta^{a \: a_{1} \ldots  a_{k}}(\tau)  \nabla_{a} \nabla_{a_{1}} \ldots \nabla_{a_{k}}$ should be a Casimir operator. As the derivatives $\nabla_{a}$ build a representation of the Lie algebra of the gauge group, this means that each such monomial should commute with $\nabla_{a}$:
\begin{equation}
\label{CasimirsDefinition}
\left[ \nabla_{b}, \eta^{a \: a_{1} \ldots  a_{k}}(\tau)  \nabla_{a} \nabla_{a_{1}} \ldots \nabla_{a_{k}} \right] = 0
\end{equation}
Examples of such monomials are $\Delta^{2} = \delta^{ab} \delta^{cd} \nabla_{a} \nabla_{b} \nabla_{c} \nabla_{d}, C_{adj} \Delta = C_{ab}^{c} \nabla_{a}  \nabla_{b}  \nabla_{c}$ and so on. It is now straightforward to show that if the relation (\ref{CasimirsDefinition}) holds for all differential operators on the r.h.s. of (\ref{KramersMoyall}), the solution of the equation (\ref{KramersMoyall}) will be the function on group classes. Indeed, for $\tau = 0$ $p(g, \tau) = \delta(g, 1)$ is the function on group classes. From the commutation relations (\ref{LeftRightCommutationRelations}) it follows that the derivatives $\tilde{\nabla}_{a}$ commute with arbitrary functions of the derivatives $\nabla_{a}$, and hence $\left[\nabla_{b} - \tilde{\nabla}_{b}, \eta^{a \: a_{1} \ldots  a_{k}}(\tau)  \nabla_{a} \nabla_{a_{1}} \ldots \nabla_{a_{k}} \right] = 0$ whenever (\ref{CasimirsDefinition}) holds. This implies that if $\left( \nabla_{b} - \tilde{\nabla}_{b} \right) p(g, 0) = 0$, $\left( \nabla_{b} - \tilde{\nabla}_{b} \right) p(g, \tau) = 0$ for all $\tau$. Thus gauge invariance of the cumulants (\ref{LieAlgebraCumulants}) guarantees that the solution of the equation (\ref{KramersMoyall}) always belongs to the Hilbert space of functions on group classes.

 Gauge invariance requires that the "drift term" $\eta^{a}$ should vanish and that $\eta^{a b}(\tau) = C(\tau) \delta^{a b}$, where $C(\tau)$ is some function of $\tau$ which is proportional to the second-order cumulant of the shifted curvature tensor. In the approximation of a gaussian-dominated QCD vacuum one assumes that the only non-zero cumulant is the second-order cumulant \cite{Dosch:02, Simonov:96, Steffen:03}. In this case the equation (\ref{KramersMoyall}) reduces to the well-known heat kernel equation, in agreement with the results of \cite{Brzoska:05, Buividovich:06:2, Arcioni:05}:
\begin{eqnarray}
\label{GaussianDiffusion}
\partial_{\tau} p(g, \tau) = C(\tau) \Delta p(g, \tau)
\end{eqnarray}
Such limit of the equation (\ref{GeneralFinalEquation}) automatically leads to Casimir scaling (which is not surprising, since the gaussian dominance conjecture was devised to explain Casimir scaling), while possible deviations from the Wilson area law can be incorporated in the function $C(\tau)$.

 General form of the equation (\ref{GeneralFinalEquation}) suggests that decay rates of Wilson loops in different representations of the gauge group should be proportional to some combination of Casimir operators, in full accordance with the analysis based on the method of field correlators \cite{Dosch:02, Simonov:96, Steffen:03}. Note that such conclusion is in general not true, for instance, if the "drift term" $\eta^{a}$ is not equal to zero, as was proposed in the works \cite{Brzoska:05, Arcioni:05}.

\section{Conclusions}
\label{sec:Conclusions}

 In this paper we have derived the differential equation for the probability distribution of parallel transporters on the gauge group. It turned out that it is necessary to add the terms with higher-order derivatives to the simplest heat kernel equation  \cite{Brzoska:05, Buividovich:06:2, Arcioni:05} in order to reproduce correctly deviations from Casimir scaling and to describe the screening effects correctly. Gauge invariance requires that all the differential operators in the r.h.s. of the equation (\ref{GeneralFinalEquation}) should be Casimir operators. In this case the tension of QCD string is in general proportional to some combination of Casimir operators, as the method of field correlators predicts \cite{Dosch:02, Simonov:96, Steffen:03}. In contrast to the phenomenological approach of \cite{Brzoska:05, Buividovich:06:2, Arcioni:05} the equation (\ref{GeneralFinalEquation}) contains no arbitrary potential functions and is thus explicitly invariant under group transformations. Moreover, the coefficients $\eta^{a \: a_{1} \ldots  a_{k}}(\tau)$ are directly related to nonperturbative cumulants of the shifted curvature tensor and hence the equation (\ref{GeneralFinalEquation}) contains no phenomenological parameters at all and is exact as long as the cumulant expansion converges.

 The final equation (\ref{GeneralFinalEquation}) is the Kramers-Moyall cumulant expansion, which generalizes the Fokker-Planck equation onto arbitrary random walks \cite{Brzoska:05, Arcioni:05, RiskenFokkerPlanck}. Such expansion arises naturally when one considers an arbitrary discrete random walk and then takes the continuum limit \cite{RiskenFokkerPlanck}. In lattice theories the set of loops is a discrete set, and the initial equation (\ref{PolyakovEquationOneParametric}) turns into an equation with discrete variable $\tau$, which corresponds to a discrete random walk on the group manifold, according to the interpretation of \cite{Brzoska:05, Arcioni:05}. One can note an interesting similarity between the continuum limit of discrete random walks on Lie groups and the continuum limit of lattice gauge theories. In particular, in both cases continuum limit means that physical observables do not depend on time step or lattice spacing.

 As long as the screening effects are small, the higher-order differential operators in (\ref{KramersMoyall}) can be treated as a small perturbation of the free diffusion described by the equation (\ref{GaussianDiffusion}). As it is known that the equation (\ref{GaussianDiffusion}) is solved by pure two-dimensional Yang-Mills theory living on the surface spanned on the loop $\gamma(\tau)$ \cite{Buividovich:06:2}, it can be particularly interesting to find a small perturbation of two-dimensional Yang-Mills theory which allows to solve the general equation (\ref{GeneralFinalEquation}) at least up to several orders in the perturbation strength. In the case of non-zero "drift term" $\eta^{a}$ such modification was found in \cite{Arcioni:05}. Indeed, the dimensional reduction scenario relates the observables of four-dimensional and two-dimensional theories \cite{Makeenko:84}, which automatically leads to Casimir scaling. Thus possible modifications of two-dimensional theory can be used to extend the applicability of the dimensional reduction scenario. Such modifications may be also helpful in finding effective actions for QCD strings which take screening into account.

 Possible generalization of (\ref{GeneralFinalEquation}) onto sets of loops which sweep multiply connected worldsheets can be also interesting. One can also implement the methods described in the section \ref{sec:GroupGeometry} to describe diffusion of higher-order differential forms. Finally, if it was possible to obtain an equation similar to (\ref{GeneralFinalEquation}), for example, basing on the Schwinger-Dyson equations for quantum Yang-Mills fields, this will provide some information on nonperturbative cumulants of the curvature tensor.


\end{document}